# In-Situ 3D Nano-Printing of Freeform Coupling Elements for Hybrid Photonic Integration

P.-I. Dietrich[1,2,3,*], M. Blaicher[1,2], I. Reuter[1,2], M. Billah[1,2], T. Hoose[1,2], A. Hofmann[4], C. Caer[5], R. Dangel[5], B. Offrein[5], U. Troppenz[6], W. Freude[2], and C. Koos[1,2,3,**]

[1] Inst. of Microstructure Technology (IMT), Karlsruhe Inst. of Technology (KIT), Hermann-von-Helmholtz-Platz 1, 76344 Eggenstein-Leopoldshafen, Germany

[2] Inst. of Photonics and Quantum Electronics (IPQ), Karlsruhe Inst. of Technology (KIT), Engesserstr. 5, 76131 Karlsruhe, Germany

[3] Vanguard Photonics GmbH, Hermann-von-Helmholtz-Platz 1, 76344 Eggenstein-Leopoldshafen, 76227 Karlsruhe, Germany

[4] Inst. for Applied Computer Science (IAI), Karlsruhe Institute of Technology (KIT), 76131 Karlsruhe, Germany

[5] IBM Research – Zürich, Säumerstrasse 4, 8803 Rüschlikon, Switzerland,

[6] Fraunhofer Inst. for Telecommunications, Heinrich Hertz Inst. (HHI), Einsteinufer 37, 10587 Berlin, Germany

*p-i.dietrich@kit.edu, **christian.koos@kit.edu

**Hybrid photonic integration exploits complementary strengths of different material platforms, thereby offering superior performance and design flexibility in comparison to monolithic approaches. This applies in particular to multi-chip concepts, where components can be individually optimized and tested on separate dies before integration into more complex systems. The assembly of such systems, however, still represents a major challenge, requiring complex and expensive processes for high-precision alignment as well as careful adaptation of optical mode profiles. Here we show that these challenges can be overcome by in-situ nano-printing of freeform beam-shaping elements to facets of optical components. The approach is applicable to a wide variety of devices and assembly concepts and allows adaptation of vastly dissimilar mode profiles while considerably relaxing alignment tolerances to the extent that scalable, cost-effective passive assembly techniques can be used. We experimentally prove the viability of the concept by fabricating and testing a selection of beam-shaping elements at chip and fiber facets, achieving coupling efficiencies of up to 88 % between an InP laser and an optical fiber. We also demonstrate printed freeform mirrors for simultaneously adapting beam shape and propagation direction, and we explore multi-lens systems for beam expansion. The concept paves the way to automated fabrication of photonic multi-chip assemblies with unprecedented performance and versatility.**

Single-mode chip-to-chip and fiber-to-chip interfaces have been a fundamental challenge in integrated optics ever since[1]. A wide variety of approaches has been explored to achieve low-loss coupling with assembly processes of manageable technical complexity[2]. The challenge comprises two aspects: First, on-chip waveguides frequently feature small mode-field diameters, in particular when high index-contrast silicon-photonic (SiP) or InP-based components are involved. As a consequence, low-loss coupling requires highest positioning accuracy of optical components, and this can only be achieved by slow and expensive[3] active alignment techniques, where the coupling efficiency is continuously monitored while optimizing the position of the devices[2,4]. Second, fiber-chip interfaces and chip-chip-interfaces of devices realized on different material platforms often have to cope with vastly different mode profiles and emission directions, that need to be carefully adapted for efficient coupling. This requires additional elements such as micro-lenses[5], micro-mirrors[6] or waveguide-based spot-size converters[7], which results in bulky systems and increases the number of components that need to be actively aligned.

To tackle these challenges, a variety of concepts for coupling of light to optical chips has been explored. These approaches can be broadly subdivided into two basic concepts: Edge coupling to etched or polished waveguide facets, and surface coupling using grating structures that are etched into the top surface of high index-contrast waveguide cores[8-11]. Edge coupling has become the mainstay for coupling of InP-based lasers or optical amplifiers to single-mode fibers[12] (SMF) and is also widely used for SiP circuits[13]. For efficient coupling, lensed fibers[12,14,15] or discrete microlenses[5,16] are used to reduce the generally rather large mode fields of SMF to those found at the chip facet. This usually leads to tight alignment tolerances, which renders active alignment indispensable[17]. In addition, the preparation of edge-emitting waveguide facets requires delicate polishing techniques that complicate fabrication[18]. The same applies to grinding or etching of lensed fiber tips, in particular when elliptical mode fields for coupling to rectangular waveguides are required[15,19]. In addition, the numerical aperture (NA) of lensed fibers is limited such that mode field diameters of less than 2 µm are generally difficult to achieve. For edge coupling to SiP circuits, additional on-chip mode field expanders are hence required[20], featuring typical lengths of hundreds of micrometers and thus consuming precious on-chip area. The same applies to adiabatic mode converters[21], which rely on continuous transformation of mode profiles by evanescent coupling to tapered silicon photonic waveguides. Surface coupling can considerably relax the alignment tolerances to values[17] of ± 2.5 µm, but still requires active alignment[3]. In addition, grating couplers are inherently wavelength-dependent with typical bandwidths of less than 30 nm, and the resulting assemblies often lead to unfavorable arrangements with fibers mounted typically perpendicularly to the chip surface. Taken together, the manifold challenges and limitations related to chip-chip and fiber-chip coupling represent a major obstacle towards widespread application of hybrid photonic integrated circuits[1,2].

In this paper we show that these limitations can be overcome by *in-situ* printing of ultra-compact beam-shaping elements onto facets of optical devices. We exploit direct-write two-photon laser lithography[22–24], which has recently been used to realize, e.g., multi-lens objectives for endoscopic imaging[25], collimating optics[26] for light-emitting diodes (LED), as well as phase masks on fiber facets[27]. We show beam forming elements with virtually arbitrary three-dimensional geometry, which do not only allow adaptation of vastly different mode profiles, but can also relax alignment tolerances to the extent that simple and cost-efficient passive assembly processes can be used. In first a set of experiments, we expand on our earlier work[28,29] and demonstrate highly efficient coupling between SMF and edge-emitting semiconductor lasers using beam shaping elements either at the laser or the fiber facet, thereby demonstration coupling losses down to 0.6 dB (coupling efficiencies η of up to 88 %). In a second set of experiments, we further show that *in-situ* printing of curved mirror surfaces for simultaneously defining beam shape and propagation direction opens up vast design freedom for optical assemblies by allowing to flexibly combine surface-emitting and edge-emitting devices in compact arrangements. A third set of experiments is finally dedicated to relaxing alignment tolerances by using multi-lens beam expanders on both device facets. Using single-mode fibers as well-defined model system, we demonstrate coupling with 1 dB position tolerances of ± 5.5 µm, sufficient for passive alignment[6,30]. Our experiments represent the first demonstration of low-loss single-mode coupling to optical chips using *in-situ* printing of beam-shaping elements. The technique can be readily applied to wide variety of optical chips and fibers, does not require any delicate facet preparation or polishing, and renders bulky on-chip mode converters unnecessary. We expect that our approach has the potential to fundamentally change performance and economics of hybrid photonic systems.

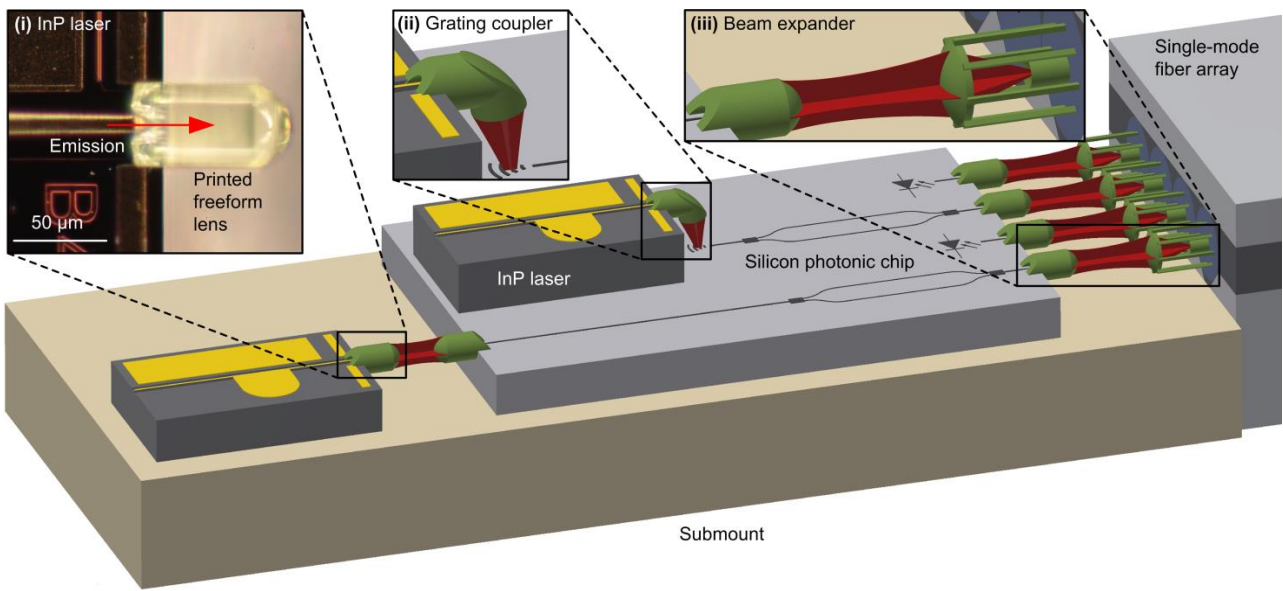

**Figure 1: Photonic multi-chip assembly** combining the distinct advantages of different photonic integration platforms: Optical sources are realized on direct-bandgap InP substrates and connected to a silicon photonic chip which comprises, e.g., Mach-Zehnder modulators and SiGe photodiodes for generation and detection of optical signals. A single-mode fiber (SMF) array is used for connection to the outside world. Freeform beam-shaping optical elements such as micro-lenses or micro-mirrors are used for mode field adaptation (red), thereby enabling low-loss coupling and relaxed alignment tolerances. Inset (i): Microscope image of a freeform lens printed on the facet of an edge-emitting laser. Inset (ii): Schematic of a freeform mirror printed on the facet of an edge-emitting laser to enable grating coupling. Inset (iii): Schematic of a pair of beam expanders.

## Concept, design, and fabrication

The concept of a hybrid photonic multi-chip assembly is illustrated in Figure 1. The assembly combines the distinct advantages of different photonic integration platforms: Lasers on direct-bandgap InP substrates are used as optical sources to deliver light to a silicon photonic (SiP) chip. The SiP is connected to the outside world via an array of single-mode fibers. Within the assembly, chip-chip and fiber-chip coupling is accomplished by beam-shaping elements, which are fabricated *in-situ* at the facets of the associated optical components using 3D two-photon lithography, see Methods Section for details about simulation and fabrication. Inset (i) shows a microscope image of a freeform lens printed to the facet of an edge-emitting laser. By exploiting high-resolution machine vision techniques, the beam-shaping elements can be aligned to the respective waveguides with precision of better than 100 nm. This capability along with the virtually unlimited design freedom of printed freeform structures allows to precisely match the vastly different mode field profiles of the various components and hence to achieve low-loss coupling. Printed reflective elements with precisely defined curved mirror surfaces allow to flexibly match the propagation directions of optical beams, thereby interfacing, e.g., edge-emitting lasers to grating couplers on the chip surface, see Inset (ii). In this assembly, alignment tolerances can be greatly relaxed by transforming small mode fields at the device facets into collimated beams with large diameters, Inset (iii). This allows assembly of multi-chip systems with relaxed alignment tolerances of ± 5 µm or above, such that high-throughput passive alignment techniques can be used.

In our experiments, we explore and demonstrate a toolbox of essential beam-shaping elements that can be used as universal building blocks for hybrid multi-chip systems. The elements are illustrated in Figure 2 and in the associated insets. The corresponding scanning electron microscope (SEM) images of printed elements are depicted in Figure 2 (a)−(f). The beam shaping elements can be realized as freeform lenses having a single refractive surface, Figure 2 (A,a) and (C,c), or as reflecting elements exploiting freeform mirror surfaces, Figure 2 (B,b) and (D,d). In addition, more complex structures are possible, comprising, e.g., combinations of concave and convex lenses for expanding beam diameters, Figure 2 (E,e), or high-performance multi-lens assemblies, Figure 2 (F,f). The beam-shaping elements can be designed for operation in air or in a low-index cladding material, which reduces reflection and protects the optical surfaces from environmental influences. Note that the cladding material also reduces the refractive power of an individual optical interface, which must be compensated by using a multitude of cascaded surfaces in a multi-lens assembly, Figure 2 (F,f), see also Section IV of the Supplementary Information. Details of the fabrication process can be found in the Methods Section. By an appropriate choice of lithography parameters, surfaces with optical quality and a root-mean-square roughness of 37 nm are attained, which can be further decreased using adapted writing strategies, see Figure S1 in the Supplementary Information for more details. The printed components withstand high powers up to 3 W and maintain constant coupling efficiency over an optical bandwidth of more than 100 nm, see the Methods Section for more details.

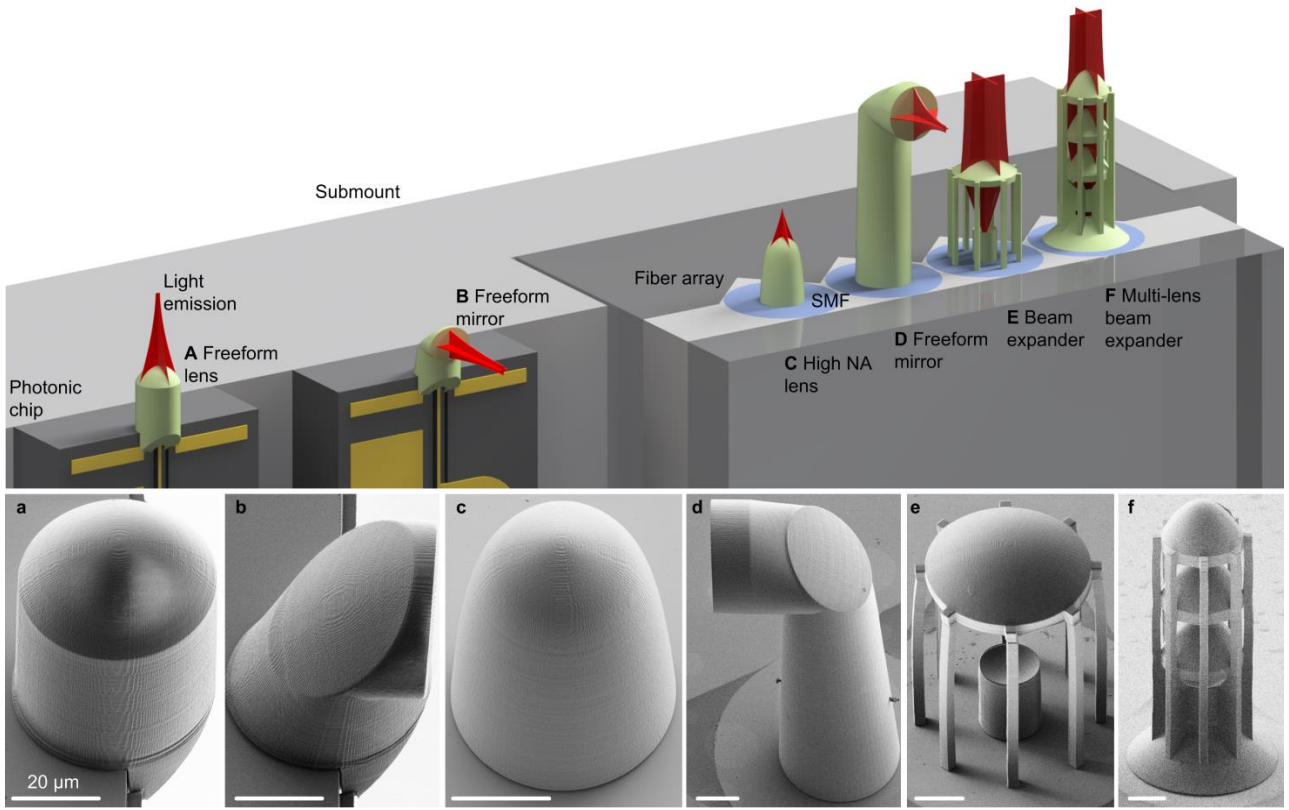

**Figure 2: Artist's view (A … F) and experimental realizations (subfigures a … f) of various beam-shaping elements that can be used as universal building blocks for hybrid photonic multi-chip systems. (A,a)** Freeform lens and **(B,b)** total internal reflection mirror printed to the facet of an edge-emitting laser; **(C,c)** Freeform lens with high numerical aperture (NA), **(D,d)** total internal reflection mirror for beam deflection, all printed to a the facet of a SMF to couple light from an surface-emitting or edge-emitting lasers. **(E,e)** Beam expander, **(F,f)** multi-lens optics featuring nine refractive surfaces, all designed for relaxing alignment tolerances of the assembly process.

## Experimental verification and discussion

The basic toolbox of beam-shaping elements contains three essential building blocks: Facet-attached lenses for mode-field adaption in edge-coupled interfaces, free-form mirrors that allow to combine surface and edge-emitting devices in compact arrangements, as well as beam expanders that relax positioning tolerances in the assembly process. In the following we describe our experimental validation of these structures.

**Coupling experiments with facet-attached lenses.** In a first set of experiments, we demonstrate coupling of edge-emitting distributed feedback (DFB) lasers to SMF. For such assemblies, beam-shaping elements can be printed either to the laser or the SMF facet, see Figure 3 (a, d). In both cases, we determined the coupling efficiency as well as its sensitivity with respect to misalignment. To this end, we moved the SMF in horizontal ($x$), vertical ($y$) and axial ($z$) direction while measuring the fiber-coupled power, see Figure 3 (b, c) and (e, f). For beam-shaping lenses attached to the laser facet, the coupling loss amounts to 1.0 dB ($\eta = 80\,\%$) for the optimum position of the fiber. This is close to the theoretical optimum of 0.6 dB that can be expected for this configuration, see Methods for details of the simulation and the measurement techniques. An SEM image of a printed lens on laser facet is shown in Figure 2 (a). Facet-attached beam-shaping lenses are particularly well suited for coupling to ultra-small mode-field diameters by exploiting the principle of solid-immersion lenses, which allow the reduction of beam divergence by the higher refractive index of the lens medium. Still, the alignment tolerances are dictated by the rather large mode-field diameter (MFD) of the SMF and the associated beam divergence. In our experiment, we find a horizontal and vertical 1 dB positioning tolerance of ± 1.9 µm, little below the simulated tolerance of ±2.4 µm, indicated by the dashed line in Figure 3 (b). In axial direction, the 1 dB positioning tolerance amounts to ±25 µm. This result is on par with lens structures obtained by highly complex laboratory-scale ion beam milling techniques[31], which are not suited for industrial production.

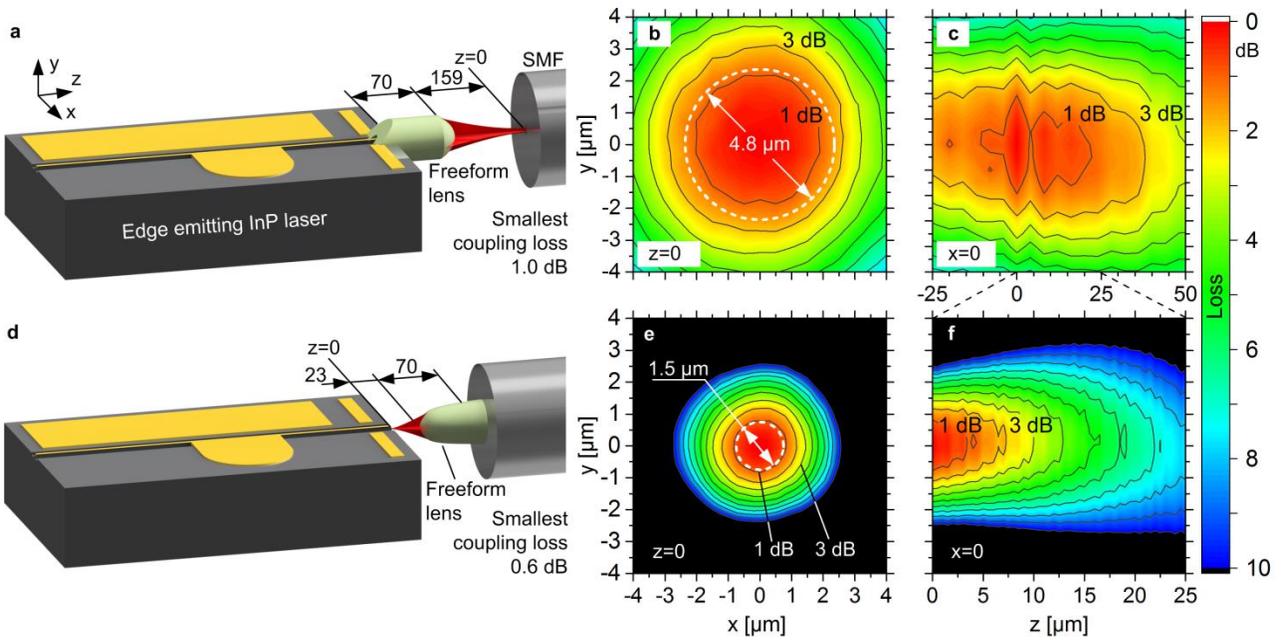

**Figure 3: Coupling of edge-emitting distributed feedback (DFB) lasers to single-mode fibers (SMF).** Beam-shaping lenses are printed either on the laser facet or on the SMF. All dimensions are indicated in µm. The 0 dB-level of the contour plots (b), (c), (e), (f) correspond to the smallest coupling loss of 1.0 dB or 0.6 dB indicated in the corresponding subfigure (a) and (d). **(a)** Freeform lens printed at the facet of a laser. The coordinate system is chosen such that optimum alignment of the fiber corresponds to the position $(x,y,z) = (0,0,0)$, where a minimum coupling loss of 1.0 dB is measured. **(b)** Increase of coupling losses relative to optimum alignment for lateral translation in the (x,y)-direction. **(c)** Increase of coupling loss relative to optimum alignment for translation in the (y,z)-direction. **(d)** Freeform lens printed at the facet of the SMF. The position $(x,y,z) = (0,0,0)$ refers again to the point of best coupling with 0.6 dB loss. **(e)** Contour plot of additional coupling loss when moving the fiber in (x,y)-direction. **(f)** Contour plot of coupling when moving the fiber in (y,z)-direction.

For beam-shaping lenses attached to the fiber facet, Figure 3 (d), positioning tolerances are dictated by the smaller MFD of the beam emerging from the laser facet. The minimum coupling losses amount to 0.6 dB ($\eta = 88$ %) for optimum alignment, which is in fair agreement to the theoretical optimum of 0.2 dB dictated by Fresnel reflections, see Methods for details. In agreement with simulations, we find a 1-dB-tolerance of ± 0.7 µm for movements in $x$- or $y$-direction, indicated by the dashed line in Figure 3 (e). In $z$-direction the 1 dB tolerance amounts to 5.8 µm and was only measured in the direction of increasing distance between the fiber and chip to avoid the risk of damaging the laser facet. The demonstrated coupling efficiency is better than the 80 % efficiency achieved with best-in-class lensed fibers[12]. Unlike conventional lensed fibers realized by melting or grinding[12,15], our approach is also applicable to a wide range of fiber-optic assemblies including also fiber arrays. An SEM image of a printed high-NA lens on a SMF facet is shown in Figure 2 (c).

**Compact assemblies with facet-attached freeform mirrors.** A second set of experiments is dedicated to facet-attached freeform mirrors with curved surfaces allow that simultaneously adapt the mode profile and the propagation direction of light. This allows to flexibly combine edge-coupled and surface-coupled devices without incurring large form factors of the overall assembly. Examples comprise, e.g., coupling of vertical-cavity surface-emitting laser (VCSEL) to optical fibers that are mounted with their axis parallel to the chip surface, or combinations of edge-emitting III-V lasers with surface-emitting grating couplers of SiP circuits, see Figure 1 (ii). Printed beam-shaping mirrors may be attached to laser facets, see Figure 2 (B,b), or to fiber end faces, see Figure 2 (D,d). We use 3D-printed total internal reflection (TIR) freeform mirrors, designed to simultaneously focus and redirect the edge-emitted light into the desired direction. The viability of TIR mirrors is first tested by coupling of two SMF over an angle of 90°, see Figure 4 (a). In this experiment, we measure losses as low as 0.6 dB ($\eta = 87\%$). A SEM image of the TIR mirror element is shown in Figure 4 (b). This approach was then transferred to fiber-chip coupling, Figure 4 (c) and (d). For freeform mirror elements printed to the laser facet, Figure 4 (c), we demonstrate coupling losses of 2.9 dB ($\eta = 51\%$) compared to the total power emitted by the laser before attaching the beam-shaping element. This loss is mainly caused by the large divergence of the beam emerging from the laser facet, which leads to incomplete TIR at the mirror surface, see Methods for details. This can be overcome by optimized single TIR surfaces, possibly in combination with refractive surfaces for beam shaping, by cascaded TIR surfaces, by resist materials with higher refractive index, or by reflective coatings. Nevertheless, the demonstrated coupling losses are already now significantly lower than those obtained for similar structures such as horizontal-cavity surface-emitting lasers[32] (HCSEL) that may be complemented by monolithically integrated lenses[33] to reduce laser-to-fiber coupling losses to 4.9 dB. Alternatively, mirror elements can be attached to the facet of a SMF, e.g., for coupling of light to and from a VCSEL, see Figure 4 (d) for the concept and Figure 2 (d) for a fabricated device. With this approach, we could achieve coupling losses of 1.1 dB ($\eta = 78\%$). This is better than typical coupling losses of packaged VCSEL[34], for which values of 3 dB have been reported. In addition, our approach allows to interface low-cost surface-coupled lasers or photodetectors to SMF arrays while maintaining compact and flat form factors. The concept can be transferred to grating-coupler interfaces of SiP circuits, and should be able to outperform angle-polished fiber arrays[10] for which coupling efficiencies of 4.5 dB have been demonstrated. Moreover, facet-attached freeform mirrors lend themselves to connecting edge-emitting lasers to grating couplers[6], thereby replacing rather complex micro-optical bench (MOB) assemblies of ball lenses and micro-prisms that are currently used in commercial silicon photonic products[35]..

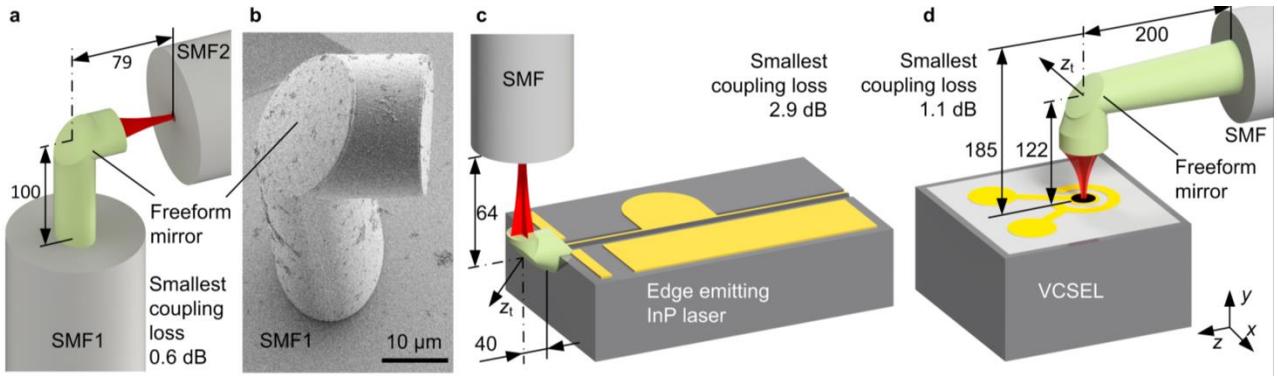

**Figure 4: Coupling experiments of optical components equipped with freeform mirrors. (a)** Model system combining a freeform mirror on the facet of a single-mode fiber (SMF) designed for coupling couple into a second SMF. Coupling losses as low as 0.6 dB are achieved. **(b)** Electron-microscopy image corresponding printed beam-shaping element. **(c)** Freeform mirror on the device facet of an edge-emitting laser designed to couple in a single-mode fiber (SMF), see Figure 2 (b,B) for the corresponding realizations of the beam-shaping element. In the experiment, we measure coupling losses of 2.9 dB, see Methods for details. **(d)** Coupling from a vertical-cavity surface emitting laser (VCSEL) to a SMF with a freeform mirror fabricated on the SMF-facet, Figure 2 (d,D). This approach allows to considerabley reduce the assembly height to less than 200 µm.

**Coupling experiments with facet-attached optical beam-expander.** The first two sets of experiments relied on single beam-shaping elements. In these cases, the alignment tolerance is inherently dictated by the MFD of the component not equipped with a beam-shaping element. To relax tolerances to a level suitable for passive alignment, beam shaping elements on both device facets can be used to expand and refocus the beam – see Figure 2 (E,e) for concepts and SEM images of fabricated elements. This is demonstrated in a third set of experiments by using a pair of SMF as a well-defined model system, see Figure 5 (a). The beam expansion allows to trade translational alignment tolerances in *x*, *y*, and *z*-direction against rotational alignment tolerances of the optical axes, which are generally easier to meet[36]. In our experiment, we demonstrate coupling losses of 1.9 dB for perfect position and tilt alignment of the fibers, see Supplementary Information for details on the measurement technique. In agreement with simulations, the 1 dB translational alignment tolerances amount to ± 5.5 µm in the transverse and to ± 220 µm in the axial direction, see Figure 5 (b) and (c). Based on these results, we estimate angular 1 dB alignment tolerances of ± 1.5°. These requirements are well compatible with passive alignment techniques[6,30]. Out of the residual loss of 1.9 dB, 1.3 dB are caused by Fresnel reflection at the six air-polymer interfaces according to simulation. The remaining 0.6 dB are attributed to imperfect fabrication and surface roughness, which may be further mitigated by an improved design and refined writing strategies of the beam expander.

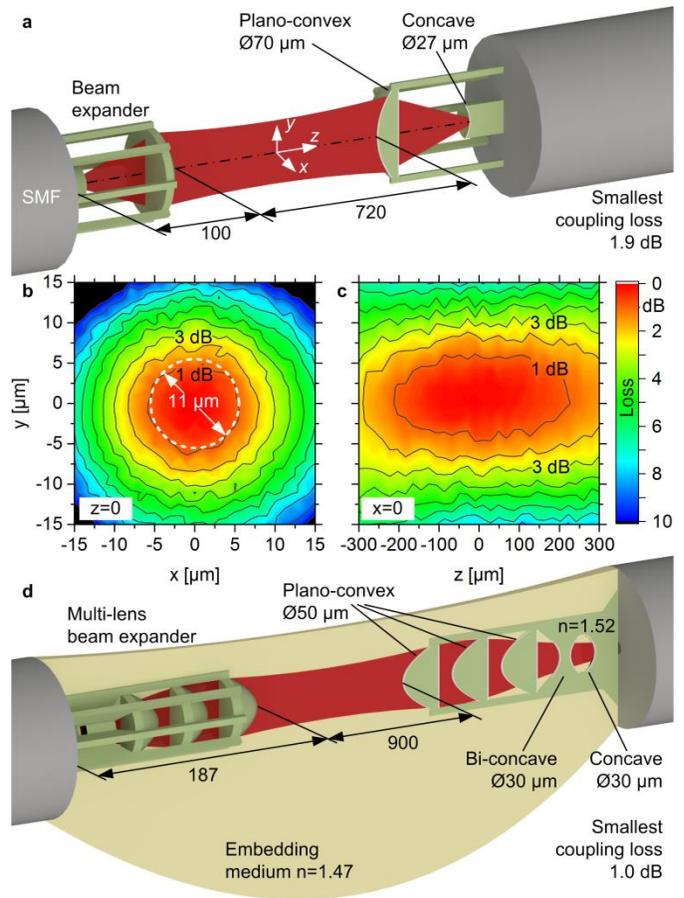

**Figure 5: Coupling experiments using expanders for relaxing alignment tolerances.** As a well-defined model system, we use single-mode fibers (SMF). **(a)** Concept for the coupling experiment using a pair of two-lens beam expanders in air. **(b)** Increase of coupling losses relative to optimum alignment for translation in the (*x,y*)-direction. A dashed circle shows simulated 1dB coupling tolerances. **(c)** Increase of coupling losses relative to optimum alignment for translation in the in (*y,z*)-direction. **(d)** Coupling experiment demonstrating a low index-contrast micro-optical system (LIMOS) that allows to significantly reduce reflection losses in printed multi-lens assmeblies.

In general, reflection at a single dielectric interface can be reduced by using a smaller refractive index contrast. This, however, reduces also the refractive power and needs to be compensated by cascading a multitude of interfaces. Multi-lens assemblies with low index contrast can be used to overcome the problem of reflection losses: In Section IV of the Supplementary Information, we show that, for a given refractive power of the overall lens system, the total Fresnel losses can be reduced by increasing the number of lens surfaces while decreasing the index contrast at each surface. We refer to such assemblies as "low index-contrast micro-optical system" (LIMOS). An implementation of the LIMOS concept is shown in Figure 5 (d), where a pair of multi-lens beam expanders of refractive index n = 1.52 is embedded into a cladding with a refractive index of $n_0 = 1.47$. An SEM image of a LIMOS beam expander is shown in Figure 2 (f).

For experimental verification, the structure is embedded into index-matching liquid that emulates the cladding. The simulated fiber-to-fiber coupling loss of the LIMOS-expander amounts to 0.1 dB ($\eta$ = 98 %) including reflection. In the experiment, this loss increases to 1.0 dB ($\eta$ = 80 %), see Supplementary Information for details about the measurement. The difference of calculated and measured losses is attributed to fabrication inaccuracy. In particular, a tilt of 0.5° of the expander axis with respect to the fiber axis causes an additional loss of about 0.6 dB. In the future, this can be mitigated by an improved tilt detection and correction. Nevertheless, the experimentally demonstrated coupling loss of the 18-surface LIMOS expander is smaller than the 1.3 dB that can be theoretically achieved by a comparable six-surface structure in air. The LIMOS approach addresses one of the most stringent challenges of printed micro-optics: Since optical surfaces in advanced printed 3D assemblies are not directly accessible, anti-reflection coating, e.g., by gas-phase deposition, is not possible. Sub-wavelength surface structures are effective in reducing reflection[37] but produce scattering loss. The LIMOS concept overcomes these challenges and we expect that this approach is applicable to a wide range of printed microlenses for both imaging and beam-shaping applications. As a side effect, the low-index cladding acts as a protective layer against environmental influences.

## Summary and outlook

We have demonstrated that 3D freeform-optical elements printed to device facets can be used for highly efficient and position-tolerant coupling. We explore and demonstrate a toolbox of essential beam-shaping elements required to build hybrid optical assemblies: Freeform lenses, freeform mirrors and multi-lens beam expanders that reduced tolerances such that passive alignment can be used. In a first set of experiments we demonstrate single freeform lenses printed either to laser or to fiber facets, leading to coupling losses of down to 0.6 dB. A second set of experiments is dedicated to freeform total-internal reflection (TIR) mirrors, that allow to combine surface- and edge-emitting optical components in compact assemblies. We demonstrate the viability of the approach by coupling vertical-cavity surface-emitting lasers (VCSEL) to single-mode fibers (SMF) using printed freeform mirrors on either the laser or the SMF end face. This lead to coupling losses of down to 1.1 dB. A third set of experiments demonstrates that beam expanders on both facets of a coupling interface allow to considerably relax translational positioning tolerances. Using a pair of SMF as a well-defined model system, we demonstrate coupling losses of 1.9 dB, mainly limited by Fresnel reflection. The lateral 1 dB alignment tolerances amount to ± 5.5 µm, which is compatible with high-throughput passive assembly techniques. This approach can be transferred to a wide variety of edge-emitting and surface-emitting devices and could pave the path towards large-scale passive alignment of optical components. For reducing the Fresnel reflection loss in multi-lens systems, we introduce the concept of low index-contrast micro-optical systems (LIMOS). The viability of the approach is shown by coupling a pair of SMF via a pair of LIMOS beam expanders that comprise a total of 18 lens surfaces. This significantly reduced coupling loss to 1.0 dB, which is even below the minimum losses that can be theoretically achieved by lenses operated in air. We expect that the LIMOS technology enables new types of printed multi-lens systems with almost no reflection loss.

## Acknowledgements

This work was supported by the BMBF Project PHOIBOS (Grant 13N12574), by the Helmholtz International Research School for Teratronics (HIRST), by the European Research Council (ERC Starting Grant 'EnTeraPIC', # 280145), by the H2020 Photonic Packaging Pilot Line PIXAPP (# 731954), by the EU-FP7 project BigPipes, by the Alfried Krupp von Bohlen und Halbach Foundation, by the Karlsruhe Nano-Micro Facility (KNMF), by the Deutsche Forschungsgemeinschaft (DFG, 1173), and by the IBM PhD Fellowship Program (P.-I. D.). We thank Philipp Trocha for help with the high-power measurements, Marco Hummel for fabricating mechanical setups, Oswald Speck for fiber preparation, Florian Rupp and Paul Abaffy for recording SEM images, and Gerald Göring and Norbert Schneider for the AFM measurements.

## Author Contributions

P.-I. D. designed, simulated, fabricated and characterized coupling structures and devices with help from M. Bl., I. R., M. Bi., T. H., and A. H., supervised by C. K.. M. Bl. supplied advanced tools and techniques for 3D printing. M. Bi. and T. H. supported fabrication and measurement of test structures. C. C., R. D. and B. O. contributed to fabrication of test chips elements. U. T. contributed InP-based components. Device concepts and coupling schemes were jointly conceived by P.-I. D., M. Bl., R. D., B. O., and C. K.. All authors discussed the data. The project was supervised by W. F. and C. K. The manuscript was written P.-I. D., W. F., and C. K..

## Additional Information

Supplementary Information is available online. Correspondence and requests for materials should be addressed to C. K.

## Competing financial interests

P.-I. D. and C. K. declare being co-founders and shareholders of Vanguard Photonics GmbH, a startup company engaged in exploiting 3D nano-printing in the field of photonic integration and assembly. P.-I. D., M. B., I. R., and C. K. are co-inventor of patents owned by Karlsruhe Institute of Technology (KIT) in the technical field of the publication.

## Methods

**Simulation:** We optimize optical freeform components by simulation with the physical-optics module of the commercial design software ZEMAX[38]. As an input for the simulation of the various components, we assume mode fields with rotationally symmetric Gaussian intensity distributions. All mode-field diameters (MFD) in the manuscript refer to the $1/e^2$-width of the intensity distribution at a wavelength of 1550 nm. The MFD of the SMF was 10 µm, while the MFD of the edge-emitting laser and the VCSEL were 3.0 µm and 7.5 µm, respectively. For optimization of the lenses, we represent the freeform shape of the lens surface by a rotationally symmetric polynomial with a height $z$ as a function of the radius $r = \sqrt{x^2 + x^2}$ of the form $z(r) = a_0 + a_2 r^2 + a_4 r^4 + a_6 r^6 + a_8 r^8$, see Figure 3 (a) for the definition of the coordinate system. For optimization of the lens surface, we neglect Fresnel reflection as it does not significantly alter the optimal shape while slowing down simulation. The surface is optimized by varying the parameters $a_0 \ldots a_8$ to achieve maximum coupling efficiency for the various configurations.

The surfaces of the reflective elements are defined in a rectangular coordinate system $x_t, y_t, z_t$, where the $z_t$-axis is tilted by an angle of $45°$ with respect to the optical-axis, see Figure 4 (c) and (d). The surface of each total internal reflection (TIR) mirror is then represented by a polynomial $z_t(x_t, y_t) = \sum_{i=0}^{4}\left(a_{2i}x_t^{2i} + b_{2i}y_t^{2i}\right)$. For simplicity, perfect reflection was assumed for optimization, i.e., the surface was treated as an ideal specular reflector irrespective of the angle of incidence. The surface is again optimized by varying the parameters $a_0...a_8$ and $b_0...b_8$, to achieve maximum coupling efficiency for the various configurations.

**Fabrication:** We print all structures by using the commercial 3D two-photon lithography system Nanoscribe Photonic Professional*GT*, equipped with a $40\times$ objective lens (numerical aperture $1.4$) as well as galvanometer mirrors for rapid beam movement in the lateral directions. The system was operated with a dedicated software, optimized for high shape fidelity of the printed beam-shaping elements and high-precision automated alignment with respect to the device facets. We use a liquid photoresist (Nanoscribe IP-Dip[39], refractive index $n = 1.52$) that simultaneously acts as an immersion medium.

**Coupling experiments with facet-attached lenses:** To determine the coupling efficiency from an edge-emitting InP laser (1590 nm, 10.8 mW at 50 mA) to a SMF, we first measure the emission power of the bare laser with an integrating sphere. Where applicable, this measurement was repeated after writing the freeform lens to the laser facet. The laser was equipped with a tapered waveguide and an anti-reflection coating at the facet. The fiber-coupled power was measured using the same integrating sphere as for the measurement of the laser power. The coupling efficiency was determined from the ratio of the two values taking into account the loss 0.16 dB due to extra Fresnel reflection loss at the plane SMF end facet pointing into the integrating sphere. Alignment tolerance measurements were performed by moving the SMF by steps of 0.5 μm in the horizontal (x) and vertical (y) direction. In axial direction (*z*), a step size of 4 μm was chosen.

The measured efficiencies were benchmarked to simulations of the respective coupling interface performed by the ZEMAX physical-optics module software[38]. For beam-shaping lenses attached to the laser facet, a loss contribution of 0.1 dB is caused by residual mode-field mismatch, whereas Fresnel reflection at the SMF-air interface contributes an additional 0.2 dB. Moreover, the laser power decreased by 0.4 dB after fabrication of the lens, which can be attributed to reflections at the laser-lens and the lens-air interfaces, see Section III of the Supplementary Information for a more detailed discussion. Assuming an ideal lens shape, the best achievable coupling losses for the current devices should hence amount to 0.6 dB. For beam-shaping lenses attached to the fiber facet, the lowest theoretically achievable coupling losses amount to 0.2 dB as dictated by Fresnel reflection at the lens-air facet. Note that the refractive indices of the SMF and the lens are nearly equal such that this interface does not contribute any additional losses. Material absorption within the lenses can be safely neglected – the interaction length within the beam-shaping elements amount to typically a few tens of micrometers, which is orders of magnitude below the absorption length of IP-Dip of typically 2.5 cm at a wavelength of 1550 nm.

**Coupling experiments with facet-attached freeform mirrors:** For coupling experiments with facet-attached mirrors, we use again the same integrating sphere to measure the free-space and fiber-coupled power levels. Coupling efficiencies are then obtained from the respective power levels by taking into account the extra Fresnel loss of 0.16 dB occuring at the plane SMF end facet pointing into the integrating sphere. The edge emitting InP laser is again operated at a wavelength of 1590 nm and a free-space output power of 10.8 mW (pump current 50 mA). For the 1550 nm VCSEL, the total output power amounts to 1.0 mW (pump current 17.6 mA).

For the solidified photoresist, we assume the refractive index of n = 1.52 that is specified for the liquid material. This leads to a critical angle of TIR of approximately 41°, measured with respect to the surface normal. For the coupling experiment using a TIR element attached to the edge-emitting InP laser, the large beam divergence may lead to configurations where a certain portion of the emitted light hits the mirror surface at angle below the critical angle for TIR. This is observed experimentally by measuring the power emitted in the direction of the laser cavity. Incomplete TIR is the main source of loss for such assemblies. This can be overcome by optimized single TIR surfaces, possibly in combination with refractive surfaces for beam shaping, by cascaded TIR surfaces, by resist materials with higher refractive index, or by reflective coatings.

**Coupling experiments with facet-attached optical beam expanders:** To determine the coupling efficiency achieved with beam expanders in air, we first perform a back-to-back measurement using a fiber connector. We then replace this direct connection by two fibers, equipped with identical beam expanders, and measure the coupling loss by optimizing the position of the two SMF. Note that this technique may lead to a slight overestimation of the fiber-to-fiber coupling losses, since the coupling experiment features one additional fiber connector compared to the back-to-back reference. The alignment tolerance was measured by moving one fiber on a pre-defined grid by steps of 0.5 µm (4 µm) in the lateral (axial) direction. The MFD between the two expanders operated in air was designed to a MFD of 23 µm, leading to a theoretically predicted 1-dB coupling tolerance of ±5.5 µm, which is confirmed by experiments.

The coupling experiment was repeated with the expanders designed for operation in an embedding medium. We drop-casted index matching oil with a refractive index $n_0 = 1.47$ onto the coupling interface. Since the oil meniscus between the two SMF was unstable, it was not possible to perform reliable measurements of coupling tolerances, that require time scales up to one hour. The MFD between the expanders operated in an embedding medium were designed to a MFD of 22 µm, slightly smaller than that of the expanders operated in air.

**Surface roughness and optical performance:** The root-mean-square (RMS) surface roughness of a typical lens is investigated by atomic force microscopy (AFM). To this end we used a lens printed on a laser facet, see Figure 2 (A,a). Within a $25 \times 25$ µm² area we measure a RMS surface roughness of 37 nm, see Supplementary Information Figure S1 (a). Coupling experiments at a wavelength of 1550 nm indicate that the measured lens roughness does not degrade the coupling efficiency to any significant amount, see Supplementary Information Section II. RMS roughness of the order of 3 nm can be reached by using optimized writing strategies or resist with smaller spatial resolution, see Supplementary Information Figure S1 (b).

**High-power damage level measurement:** To determine the damage level of beam-shaping elements for high optical powers, we use a lens on a SMF facet, see Figure S2 of the Supplementary Information. The fiber is connected to an external-cavity laser operated at 1550 nm, followed by an erbium-doped fiber amplifier (EDFA) that has a maximum output power of 36 dBm. The amplifier is protected by an isolator. Another bare SMF is used to collect the light emitted from the lens, and the coupling efficiency was monitored while increasing the output power in the range (20…36) dBm in steps of approximately 1 dB. At each level, the power was kept constant for at least one minute. At a power of 35.6 dBm (3.6 W) the coupling efficiency degraded abruptly, and the lens disintegrated in a flash of visible light.